\documentclass[twocolumn,superscriptaddress,showkeys,amsmath,amssymb,pra]{revtex4}

\usepackage{graphicx,amssymb}
\usepackage{dcolumn}
\usepackage{bm}

\newcommand{\bfr}{{\bf r}}
\newcommand{\beq}{\begin{equation}}
\newcommand{\eeq}{\end{equation}}
\newcommand{\bea}{\begin{eqnarray}}
\newcommand{\eea}{\end{eqnarray}}

\def\bfr{{\bf r}}

\begin{document}

\title{Non-equilibrium dynamics: Studies
of reflection of Bose-Einstein condensates.}
\author{R. G. Scott}
\affiliation{The Jack Dodd Centre for Photonics and Ultra-Cold Atoms, Department of Physics, University of Otago, Dunedin, New Zealand}
\author{C. W. Gardiner}
\affiliation{The Jack Dodd Centre for Photonics and Ultra-Cold Atoms, Department of Physics, University of Otago, Dunedin, New Zealand}
\author{D.~A.~W. Hutchinson}
\affiliation{The Jack Dodd Centre for Photonics and Ultra-Cold Atoms, Department of Physics, University of Otago, Dunedin, New Zealand}
\affiliation{Laboratoire Kastler Brossel, \'Ecole Normale Sup\'erieure, 24 rue Lhomond, 75231 Paris Cedex 05, France}


\begin{abstract}

The study of the non-equilibrium dynamics in Bose-Einstein condensed gases has been dominated by the zero-temperature, mean field Gross-Pitaevskii formalism. Motivated by recent experiments on the reflection of condensates from silicon surfaces, we revisit the so-called {\em classical field} description of condensate dynamics, which incorporates the effects of quantum noise and can also be generalized to include thermal effects. The noise is included in a stochastic manner through the initial conditions. We show that the inclusion of such noise is important in the quantitative description of the recent reflection experiments.

\end{abstract}

\keywords{Bose-Einstein condensation, quantum noise, Wigner function}
\maketitle

\section{Introduction}

The usefulness of the Gross-Pitaevskii equation (GPE)\cite{Dalfovo} in describing the dynamics of Bose-Einstein condensates (BECs) has been unreasonable in its effectiveness. At first glance, this non-linear Schr\"odinger equation describing the evolution of a purely classical field, corresponding to the order parameter, or wave function of the macroscopically occupied state, should only have any validity in regimes close to equilibrium at temperatures approaching absolute zero. Even in this regime, the apparent neglect of quantum fluctuations could in principle be a severe limitation. In practise, the GPE has been very effective in describing a wide array of experiments, many involving dynamical processes far from equilibrium. This success is rooted in the fact that the quasiparticle excitations of the Bose condensed system are also the collective excitations of the system\cite{Griffin}. As long as modes of excitation are highly occupied (mode occupancy large compared to unity), their evolution can be described within the GPE framework. Thus the GPE starting from a non-equilibrium configuration, to at least a certain extent, contains within it a description of the excited states and hence a representation of a thermal component. To make this more concrete and to indicate how quantum fluctuations can be included, we briefly review the {\em classical field method} and applications to BECs.

\section{The classical field method.}

Classical field methods, although suggested early in the development of quantum mechanics\cite{Wigner}, were most extensively developed in the context of quantum laser physics\cite{Gardiner,Walls}. The name is somewhat misleading in the sense that the methods were developed to account for explicitly quantum effects such as quadrature squeezing\cite{Hans} and originates from the expression of the density operator in a basis of coherent states using a classical ``quasi-probability'' distribution function. If we assume for the moment that such a distribution exists, one can show that the action of any quantum mechanical operator on the density operator is equivalent to that of a classical operator on the quasi-probability distribution function. Thus, the quantum mechanical problem, which is described by the evolution of the density operator as prescribed by a master equation, is transformed into a (classical) partial differential equation for the evolution of the distribution function. 

Two main approaches, using different quasi-probability functions, have been used in the treatment of BECs. Specifically, these have used the positive-$P$ representation\cite{Glauber,Sudarshan} and, as we shall in this paper, the Wigner function\cite{Wigner,Steel,Sinatra,Norrie_PRL}, which was originally formulated as early as 1932. The $P$-function is used to express the density operator in the form
\beq
\hat \rho(t) = \int d^2\alpha \hspace{1mm} P(\alpha,\alpha^*,t)|\alpha\rangle \langle\alpha|,
\eeq
where $|\alpha\rangle$ is a coherent state and the function $P(\alpha,\alpha^*,t)$ is interpreted as a quasi-probability distribution. The ``quasi'' in the nomenclature is due to the fact that $P$ is not necessarily positive-definite. The subclass of representations where it is are referred to as the positive-$P$ representations.

\subsection{The truncated Wigner method.}

In this work we do not use the $P$-representation, but instead the (multi-mode) Wigner function\cite{Norrie_thesis} defined as
\bea
W(\alpha_j,\alpha_j^*,t) &\equiv& \frac{1}{\pi^{2M}} \int d^2\lambda_1 ... \int d^2\lambda_M \hspace{1mm} \hspace{2mm} \times  \nonumber \\
\prod_{j=1}^M {\rm exp} [&-&\lambda_j \alpha_j^* + \lambda_j^* \alpha_j ] \chi_{\rm W}(\lambda_j,\lambda_j^*,t),
\eea
with characteristic function
\beq
\chi_{\rm W}(\lambda_j,\lambda_j^*,t)\equiv {\rm Tr}\left\{ \hat \rho(t) \hspace{1mm} \prod_{j=1}^M {\rm exp} [\lambda_j \hat a_j^\dagger - \lambda_j^* \hat a_j] \right\}
\eeq 
where $\hat a_j^\dagger$ is the usual creation operator for an individual bosonic atom in the mode $j$.

The equation of motion for the density operator
\beq
i\hbar \frac{d\hat\rho(t)}{dt}=\left[\hat H(t), \hat \rho(t) \right]
\eeq
can then be written in terms of the Wigner function through the use of certain operator equivalences. For example,
\beq
\hat a_j^\dagger \hat \rho(t) \rightarrow \left(\alpha^*_j -\frac{1}{2}\frac{\partial}{\partial \alpha^*_j}\right) W(\alpha_j,\alpha^*_j,t),
\eeq
etc.

To then proceed one considers, not the Wigner function evolution directly, which would be computationally intractable, but the stochastic differential equations which describe a single trajectory of a specific realisation of the system through phase space. Taking sufficient of these trajectories then allows the reconstruction of the Wigner function. To convert to the appropriate stochastic differential equations, it is necessary to truncate the full evolution, neglecting terms involving cubic derivatives. Fortunately this truncation can be justified provided that the total particle density is large compared to the mode density, which is the case in our simulations. In this manner a stochastic differential equation, of a form identical to the GPE, is obtained for each mode in the low energy subspace (below some cut-off) of the problem.

More heuristically, one can think of the truncated Wigner method as simulating vacuum fluctuations through the addition of appropriate classical random fluctuations to the coherent field of the BEC's initial equilibrium state. To illustrate this let us explicitly consider the mode expansion of the wavefunction $\Psi(\bfr,t)$ given by
\beq
\Psi(\bfr,t) = \frac{1}{\sqrt{V}}\sum_{j=1}^M \alpha_j(t) \hspace{1mm}{\rm exp}[i {\bf k}_j \cdot \bfr],
\eeq
where $V$ is the real space volume and $\alpha_j(t)$ now the amplitude of the mode with wavevector ${\bf k}_j$ with the wavefunction normalised to the total number of atoms $N$. The mode space is taken to be spherical with the magnitude of the cut-off wavevector sufficient 
to prevent Fourier aliasing. 

At times $t<0$ we imagine a BEC in equilibrium held in a magnetic trap centred at $(-\Delta x, 0, 0)$ with potential profile
\beq
U_{\rm trap}= \frac{m}{2}\left[\omega_x^2(x+\Delta x)^2 + \omega_y^2 y^2 + \omega_z^2 z^2\right]
\eeq
such that the equilibrium mean field state is given by solving the three-dimensional GPE
\beq
i \hbar \frac{\partial \psi}{\partial t} = \left[ -\frac{\hbar^2 \nabla^2}{2m} + U_{\rm trap} + U_0|\psi|^2\right]\psi,
\eeq
where 
\beq
U_0=\frac{4 \pi \hbar^2 a}{m}.
\eeq
Parameters for the mass, $m$ and $s$-wave scattering length $a=2.9$ nm are taken to be those for $^{23}$Na.

The quantum fluctuations are introduced by combining the {\em real} particle field $\psi(\bfr)$ with a field of {\em virtual} particles $\chi(\bfr)$ to create a total field $\Psi(\bfr,0)$ where the virtual field is defined as
\beq
\chi(\bfr)=\frac{1}{\sqrt{V}} \sum_{j=1}^M \chi_j \hspace{1mm} {\rm exp}(i {\bf k}_j \cdot \bfr).
\eeq
The complex amplitudes $\chi_j$ have a Guassian distribution with $\langle \chi_i^* \chi_j \rangle = \frac{1}{2} \delta_{ij}$ and $\langle \chi_i \chi_j \rangle = 0$, meaning that on average each mode is populated by half a virtual particle, so that the total number of virtual particles is approximately $M/2$. The truncated Wigner method is then valid provided the competing conditions, that the number of modes $M$ is large enough such that the simulations are cut-off independent and small enough such that the number of virtual particles is much less than the number of real particles, are satisfied.

This yields, as does the full derivation using the truncated Wigner representation, the stochastic differential equation
\beq
i \hbar \frac{d \alpha_j}{dt} = \frac{\hbar^2 k_j^2}{2m}\alpha_j +\frac{1}{\sqrt{V}} \int d\bfr \hspace{1mm} e^{-i{\bf k}_j \cdot \bfr} \left[U_{\rm ext} +U_0 |\Psi|^2\right] \Psi
\eeq
for each mode within the low energy subspace, with $U_{\rm ext}$ now the total external potential. The dynamical evolution of the BEC can now be obtained by solving these equations using a {\em Fourth-order Runge-Kutta in the Interaction Picture} (RK4IP) algorithm\cite{Rob}.

\section{Reflection from a barrier.}

Recent experiments have been performed to demonstrate quantum reflection of BECs from silicon surfaces\cite{Pasquini1,Pasquini2}, showing their usefulness as atom mirrors and traps. Significant disruption of the condensate was sometimes seen in the reflected condensate however, especially for dense clouds at low approach velocity. Previous work based upon the GPE\cite{Robin_PRL} was able to explain much of this behaviour based upon disruption due to the mean field non-linearity, although the reflection probability and presence of an $s$-wave scattering halo could not be explained in this framework. Here we simulate two of the experiments using the truncated Wigner scheme.

At $t=0$ the harmonic trap is displaced by $\Delta x$ along the $x$-axis so that it is now centred at $x=y=z=0$. This now accelerates the BEC towards an abrupt potential step of positive height $U$ parallel to the $xy$-plane at $x=0$. The ``impact velocity'' of the cloud with the wall is thus $v_x \approx \omega_x \Delta x$. If the kinetic energy of the atoms at the barrier ($\approx \frac{m v_x^2}{2}$) is much less than $U$ then all the atoms are reflected. If $\frac{m v_x^2}{2} \gtrsim U$ then there is finite transmission.

The purpose of this paper is to simulate these recent experiments on quantum reflection from silicon surfaces. The surface is characterised by a Casimir-Polder potential, which involves a rapid potential drop at the surface. This is impossible to model within the truncated Wigner framework, or indeed with the simple GPE in a mode representation, since the potential drop accelerates atoms to very large wavevectors which would exceed any practicable cut-off limited by computer memory. The Casimir-Polder potential is very abrupt however, with all spatial variation occurring over a length scale of less than a healing length. It can therefore be approximated by an abrupt potential drop. This of course does not obviate the problem of the acceleration of particles to high wavenumbers, however, instead of a sharp potential drop, we can employ an abrupt potential barrier which has the same transmission probability as the potential drop. Furthermore, previous theoretical work\cite{Robin_PRL}, using a different representation of the GPE, has shown the BEC dynamics of reflection to be qualitatively independent of the actual form of the barrier.

Transmitted atoms are absorbed by the wall, which is simulated using an imaginary potential in the region $x>0$ such that
\beq
U_{\rm imag}= -Cx \hspace{1cm} x>0,
\eeq
where $C$ is a positive constant. This introduces a damping term into the equations of motion for the mode amplitudes in the region $x>0$ removing atoms. This imaginary potential should only remove {\em real} atoms however, and not the noise of the {\em virtual} particles. To prevent the quantum noise being absorbed by the wall, the wall must ``emit'' virtual particles at an equivalent rate, increasing each mode amplitude $\alpha_j$ by
\beq
\Delta \alpha_j= \frac{\Delta V}{\sqrt{\hbar V}}\sum_{\nu=1}^{\nu_{\rm max}} W_\nu \hspace{1mm} e^{i {\bf k}_j \cdot \bfr_\nu},
\eeq
at each time step $\Delta t$. Here $\Delta V$ is the volume around one spatial grid point and the complex amplitude $W_\nu$ has the statistical properties $\langle W_\nu \rangle = 0$, $\langle W_\nu W_\mu \rangle = 0$, and $\langle W_\nu^* W_\mu \rangle = \frac{- \Delta t U_{\rm imag}}{\Delta V} \delta_{\mu \nu}$.

\section{Simulations}

In this section we present simulations corresponding to two recent quantum reflection experiments. The earlier experimental configuration\cite{Pasquini1} we refer to as $A$ and the later\cite{Pasquini2} as $B$.

\subsection{BEC $A$}

For BEC $A$, the total particle number $N=3 \times 10^5$ and the trapping frequencies are $\omega_x=2\pi \times 3.3$ rad s$^{-1}$, $\omega_y=2\pi \times 2.5$ rad s$^{-1}$, and $\omega_z=2\pi \times 6.5$ rad s$^{-1}$. These parameters yield a peak density of $n_0 = 2.0 \times 10^{12}$ cm$^{-3}$.

\begin{figure}[!h]
\center
\includegraphics[width=70mm]{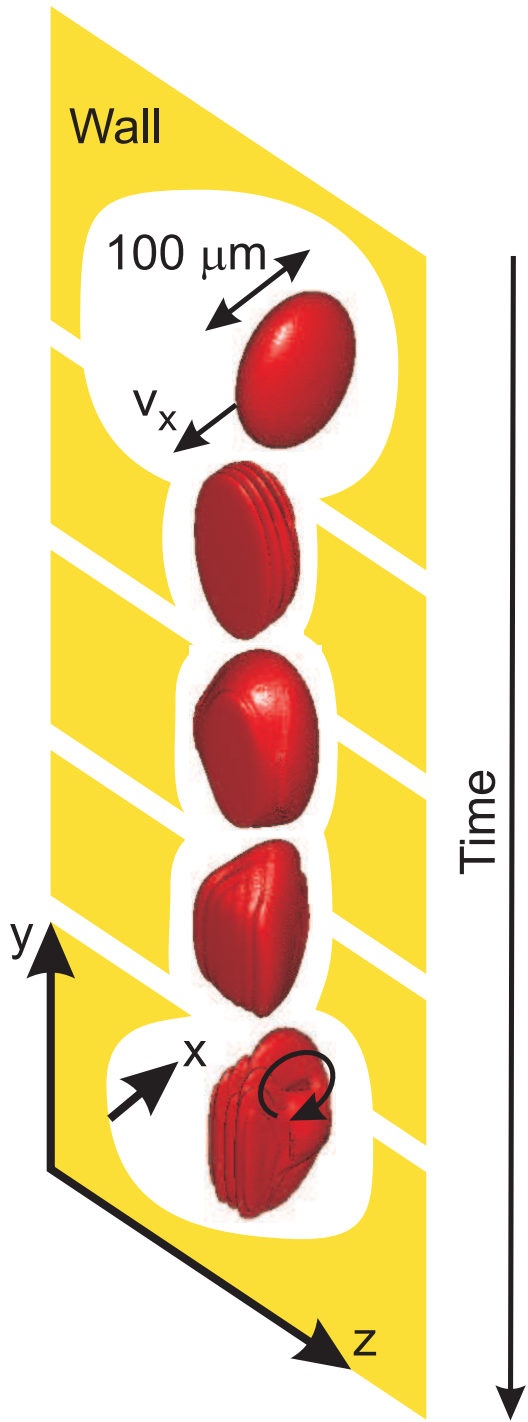}
\caption{Constant density surfaces for BEC $A$ at various times. The characteristic velocity at the barrier is 1.2 mm s$^{-1}$. The circular arrow in the final panel indicates the sense of the excited vortex state.}
\label{Fig1}
\end{figure}

Initially we perform simulations without the addition of quantum noise to investigate the role of interferential disruption in the reflection of the condensate from the potential barrier. This therefore corresponds to a simple GPE simulation. The positioning of the barrier relative to the initial trap centre ($\Delta x$) is chosen so that the characteristic velocity upon reflection is 1.2 mm s$^{-1}$. Constant density surface plots are shown in Fig.(\ref{Fig1}) for various times through the oscillation cycle back to refocusing of the BEC at the initial release point.

The density of BEC $A$ is rather low and hence the non-linearity in the GPE is weak for these simulations, corresponding to the weak mean field effects of the inter-atomic interactions. This leads to only very weak disruption of the condensate upon reflection, especially for high approach velocities when the condensate has only limited time to interact with the standing matter wave grating formed upon reflection (Fig.(\ref{Fig1}), panel 2). For characteristic velocities below about 2 mm s$^{-1}$ some disruption is seen however and vortex lines are produced in the resulting condensate as indicated in the final panel. This excitation of the internal structure of the cloud removes energy from the longitudinal motion of the condensate leading to a damping of the centre-of-mass motion\cite{Robin_PRA}. Since the excitation is due to the interference of the condensate with the standing matter wave, we refer to it as {\em interferential disruption}. The observation of such disruption is consistent with the experiments where ``excited and sometimes fragmented''\cite{Pasquini1} condensates were observed for small characteristic velocities of the cloud.

Although even in the absence of quantum noise disruption of the condensate occurs at low approach velocities, no scattering halo is produced and there is little depletion of the condensate. The depletion of the condensate comes from spontaneous processes. To simulate these we use the full truncated Wigner approach which is implemented by adding random (classical) fluctuations to the initial state as discussed in Sec.(II).

\begin{figure}[!h]
\center
\includegraphics[width=90mm]{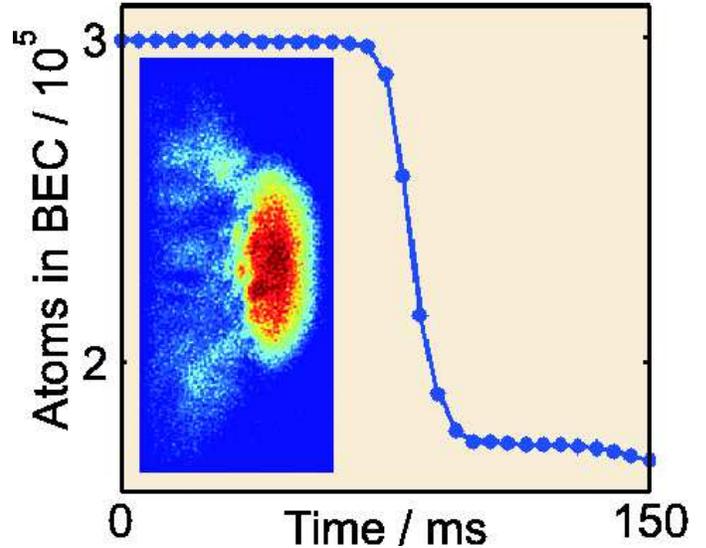}
\caption{Depletion of BEC $A$ for a characteristic approach velocity of 2.1 mm s$^{-1}$. The inset is a simulated absorption image after one complete oscillation showing the $s$-wave scattering halo.}
\label{Fig2}
\end{figure}

At low approach velocities there is little difference between the simulations with and without quantum noise. Little halo formation is seen and there is only weak depletion of the condensate. This is unsurprising as strong scattering halos were not observed in the earlier experiments\cite{Pasquini1}. By increasing the characteristic velocity (increasing $\Delta x$) however, increased depletion and the formation of a halo can be induced. This is illustrated in Fig.(\ref{Fig2}) which shows the condensate depletion as a function of time over one complete cycle for a characteristic approach velocity of 2.1 mm s$^{-1}$. The inset shows a simulated absorption image, with optical access along the $z$-direction, at the end of the completed cycle. The $s$-wave scattering halo is clearly visible corresponding to the distinct depletion of the condensate at this higher velocity.

\subsection{BEC $B$}

In the later experiments the BEC consisted of a higher total number of particles, $N= 10^6$, with a more elongated configuration having trap frequencies of $\omega_x=2\pi \times 4.2$ rad s$^{-1}$, $\omega_y=2\pi \times 5.0$ rad s$^{-1}$, and $\omega_z=2\pi \times 8.2$ rad s$^{-1}$, yielding an initial peak density of $n_0=5.2 \times 10^{12}$ cm$^{-3}$.

\begin{figure}[!h]
\center
\includegraphics[width=70mm]{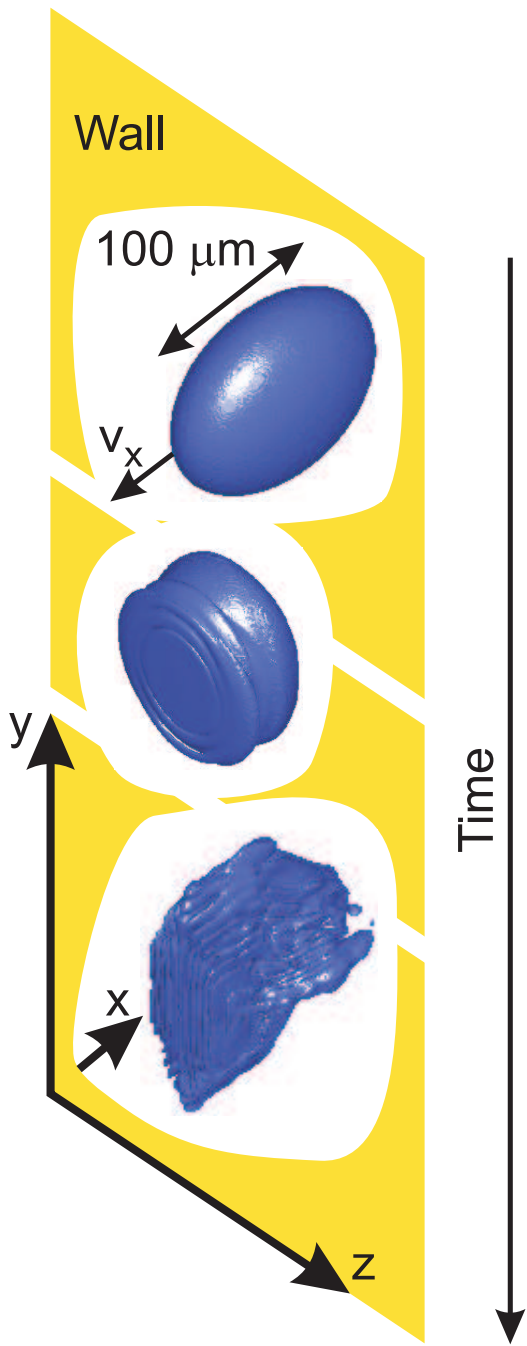}
\caption{Constant density surfaces for BEC $B$ at various times through the reflection cycle. The characteristic approach velocity is 2.1 mm s$^{-1}$. }
\label{Fig3}
\end{figure}

Since this BEC is denser and more elongated (again approaching the reflecting surface along its long principle axis), it is much more severely distorted through the interferential disruption. This is seen in Fig.(\ref{Fig3}) which displays constant density surfaces of BEC $B$ at various times through a full cycle in the absence of quantum noise. The characteristic velocity in this case is 2.1 mm s$^{-1}$ indicating that the disruption persists to significantly higher characteristic velocities. The central panel of Fig.(\ref{Fig3}) shows the formation of the standing wave. The final panel, at the end of the oscillation, shows a highly excited condensate, containing many vortices.

In the later experiments using this higher density condensate, a distinct scattering halo was observed together with a strong depletion of the condensate. This halo is absent in the simple GPE approach which was used to generate Fig.(\ref{Fig3}). Since the formation of the $s$-wave halo is a spontaneous process, quantum noise must be included. 

\begin{figure}[!h]
\center
\includegraphics[width=70mm]{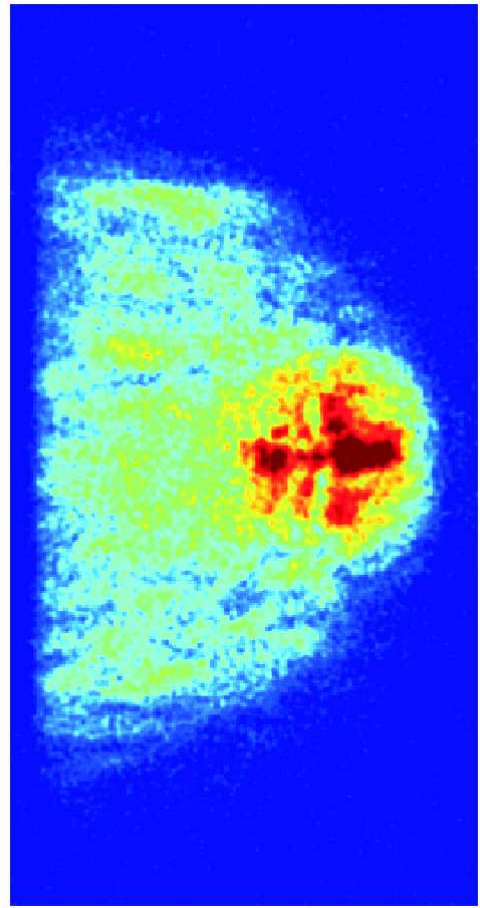}
\caption{Simulated absorption image for BEC $B$ with a characteristic approach velocity of 3 mm s$^{-1}$. The barrier height is chosen so that the reflection probability in the simulation was equivalent to that in the experiment and the image should be compared with experimental image in Fig.(4(e)) of Ref.\cite{Pasquini2}.}
\label{Fig4}
\end{figure}

Fig.(\ref{Fig4}) shows a simulated absorption image at the completion of the reflection cycle (120 ms) for condensate $B$, with a characteristic approach velocity of 3 mm s$^{-1}$ to match that of Fig.(4(e)) of Ref.\cite{Pasquini2}. The barrier height was chosen so that the reflection probability in the simulation was equivalent to that in the experiment. The scattering halo is very pronounced and the agreement between the simulated and experimental images excellent, clearly demonstrating the role of quantum noise in the reflection experiments.

\section{Conclusions}

The utilisation of abrupt barriers, such as the Casimir-Polder potential of a silicon surface, as atom mirrors has already been experimentally demonstrated\cite{Pasquini1}. The extension to more sophisticated mirror or guide geometries will require sophisticated simulation. In this paper we have developed such a fully three-dimensional technique within the truncated Wigner framework. We have noted the importance of the inclusion of quantum noise in simulating spontaneous processes such as those leading to the formation of the $s$-wave scattering halo. 

It is interesting to note that there are two competing effects leading to disruption or decoherence in the condensate after reflection. At low velocities, especially for higher density condensates, mean field effects lead to the {\em interferential disruption} of the condensate, whereas, at high characteristic approach velocities scattering into the (incoherent) $s$-wave halo has a significant depletion effect upon the BEC. This suggests that applications of BEC reflection may be cleanest for moderate approach velocities which are above the threshold for interferential disruption, but below that for the formation of the scattering halo. An alternative might be to impose a static magnetic bias field in the vicinity of the reflecting surface to turn off the interactions and hence the interference via a Feshbach resonance.

\begin{acknowledgments}
We would like to thank T. Pasquini and A. Norrie for useful discussions. This work was supported by the Marsden Fund (contracts UOO-323 and UOO-0590) and the Royal Society (London).
\end{acknowledgments}

\end{document}